Dual-comb spectroscopy with tailored spectral broadening in $Si_3N_4$ nanophotonics

Esther Baumann [1,4,*], Eli V. Hoenig[1], Edgar F. Perez[1], Gabriel M. Colacion[1], Fabrizio R. Giorgetta[1,4], Kevin C. Cossel[1], Gabriel Ycas[1,4], David R. Carlson[2], Daniel D. Hickstein[2], Kartik Srinivasan[3], Scott B. Papp[2,4], Nathan R. Newbury[1], Ian Coddington[1]

*Corresponding Author: baumann@nist.gov

[1]NIST Boulder, Applied Physics Division CO 80305, USA
[2]NIST Boulder, Time and Frequency Division CO 80305, USA
[3]NIST Center for Nanoscale Science and Technology, Gaithersburg, MD 20899, USA
[4]Department of Physics, University of Colorado Boulder, Boulder, CO 80309, USA

Abstract

$Si_3N_4$ waveguides, pumped at 1550 nm, can provide spectrally smooth, broadband light for gas spectroscopy in the important 2 µm to 2.5 µm atmospheric water window, which is only partially accessible with silica-fiber based systems. By combining $Er^+$:fiber frequency combs and supercontinuum generation in tailored $Si_3N_4$ waveguides, high signal-to-noise dual-comb spectroscopy (DCS) spanning 2 µm to 2.5 µm is demonstrated. Acquired broadband dual-comb spectra of CO and $CO_2$ agree well with database line shape models and have a spectral-signal-to-noise as high as 48/√s, showing that the high coherence between the two combs is retained in the $Si_3N_4$ supercontinuum generation. The DCS figure of merit is $6\times10^6$/√s, equivalent to that of all-fiber DCS systems in the 1.6 µm band. Based on these results, future DCS can combine fiber comb technology with $Si_3N_4$ waveguides to access new spectral windows in a robust non-laboratory platform.

1. Introduction

Frequency-comb spectroscopy can rival and exceed the signal-to-noise, speed, resolution and precision of traditional broadband spectroscopy [1–5]. Although comb spectroscopy has been shown with bandwidths over 1200 $cm^{-1}$ (~ 36 THz) in several spectral regions, [6–8] frequency-comb sources still lag behind traditional broadband thermal sources in spectral coverage. Spectral smoothness for comb systems is also a challenge as strong spectral variations can be difficult to remove from the final spectrum and may result in unusable spectral regions. While it is unlikely that any laser-based source will ever be as spectrally broad and smooth as traditional black-body sources, which include the sun, widespread adoption of comb-based spectroscopy may require the ability to generate spectrally smooth reasonably broadband light easily and flexibly across the spectrum. Moreover, to exploit coherent techniques, such as dual-comb spectroscopy (DCS), this light must maintain its coherence in both the temporal and spatial domains.

One attractive solution for generating broadband light in many spectral regions is spectral broadening in nanophotonic waveguides [9–19]. Nanophotonic waveguides have many advantages relative to traditional non-linear fibers, including compactness, high nonlinearity, control of the waveguide dispersion, broad transparency windows, and a small Raman coefficient providing high coherence [20].

One particularly promising material is silicon nitride ($Si_3N_4$, or written here as SiN), which has enabled spectral broadening in the visible, near-infrared, and mid-infrared [14–19]. SiN offers a high nonlinear index of $2.5×10^{-19}$ $m^2$/W in addition to a wide bandgap that eliminates two-photon absorption when pumped at telecommunications wavelengths [21]. Other nonlinear waveguide materials are also attractive but currently many lack the maturity of SiN, which can be fabricated into high confinement waveguides at a very high yield [22] and can also be obtained from commercial vendors.

While, recently, much of the spectral broadening in nanophotonic waveguides has been performed with an eye towards molecular spectroscopy [11–14,16,18,19] and dual-comb spectroscopy in particular, there has only been one partial waveguide based DCS demonstration [23]. This demonstration employed spectral broadening in a silicon waveguide but for only one of the two combs and with low system coherence. While this demonstration was encouraging, the coarse resolution, limited spectral coverage and low signal-to-noise leaves open questions about performance limitations.

Here we demonstrate fully coherent, high-resolution DCS with SiN waveguides. We show that DCS spectra with quality factors rivaling the state-of-the-art [3] can indeed be obtained. We also demonstrate an important caveat, which is that single spatial mode operation must be ensured. With imperfect coupler design higher order spatial modes, which are easily generated in these devices, will lead to strong and varying structure in the spectra, crippling the overall DCS performance. In this work we target the entire 2 µm to 2.5 µm atmospheric transmission window, which is of interest for space-based measurements of atmospheric gases including CO, $CO_2$, $CH_4$, $NH_3$, and $N_2O$ [24,25]. This same atmospheric window is appropriate for detection of chemicals of interest to agricultural and industrial process monitoring including HF, HCN, NH3, and acetylene. However, since this window extends beyond the transparency edge of silica, it cannot be fully accessed by broadening in nonlinear fiber optics. Previous DCS results using chromium-based solid-state lasers accessed this region but at the cost of complexity and over a narrow spectral region [26]. Accessing this spectral region through the combination of highly reliable $Er^+$:fiber combs [27–29] and SiN waveguides is an attractive option.

2. Experimental setup

Figure 1a shows a sketch of the experimental setup. Two compact fiber frequency combs with 160 MHz repetition rates (differing by 133 Hz) are phase stabilized and amplified to generate 50 fs, 1.9 nJ pulses centered at 1560 nm, with an average power of 300 mW. This amplified output is collimated and focused with two aspheric lenses onto a SiN waveguide.

The SiN waveguides were fabricated at Ligentec [30], using deep-UV lithography and chemical etching. The waveguides have a rectangular cross-section, and inverse tapers at both edge facets, which expand the mode field diameter to ~3 µm, improving input coupling efficiency of the 1560 nm $Er^+$ comb light. The SiN core is surrounded by a $SiO_2$ cladding layer as shown in the cross-section in Figure 1b. The output spectrum is tuned by choosing the waveguide dimensions and launched power appropriately [17]. To cover the 2-2.5 µm water window with high power and low spectral ripple, waveguides with a height of 770 nm, widths ranging from 1516 nm to 2020 nm and a length of 2.8 mm are chosen. All waveguides exhibit strong anomalous dispersion at the 1560 nm pump wavelength (D=75 ps/nm/km, see Figure 1c). When seeded with 1.9 nJ, these waveguides produce a supercontinuum spanning more than 120 THz while crucially providing very smooth continuous coverage throughout the desired spectral band. The total power in the generated spectra is 65 mW with 8.8 mW between 120 THz and 155 THz.

After spectral broadening in the waveguide, the light is collimated with an off-axis parabolic mirror. The output spectra are spectrally filtered by a 1.8 μm optical long-pass filter to avoid spectral aliasing and detector saturation and directed to a gas cell. The DCS signal is detected with an extended InGaAs photodiode followed by a 100 MHz transimpedance amplifier and a digitizer. The digitized time-domain interferograms are then phase corrected and co-added in real-time in a field programmable gate array (FPGA) [31]. Once a minute (every 8000 interferograms), a coadded interferogram is transferred from the FPGA to a computer and Fourier transformed to obtain a magnitude spectrum.

3. Broadening in SiN waveguides: Impact of higher-order spatial modes

Figure 2a illustrates the challenges and strengths of spectral broadening in waveguides. While spectral coverage is easily obtained, these waveguides support higher order spatial modes which can interfere to create a strong noise like structure seen in Figure 2a (green trace). Simulations suggest that these modes are excited in the waveguide, but they could also be driven by facet surface roughness or imperfect taper design. When the two combs are spatially combined these modes lead to strong inter-mode beating. Moreover, this structure varies over time and spatially across the beam front and cannot be easily removed using a reference spectrum. Fortunately, this structure can be strongly suppressed with spatial filtering after the waveguide, although, at the cost of optical power. Here we use a 10 μm diameter pinhole, which roughly matches the input beam's first Airy disc $1/e^2$-diameter of 8 μm, given by the focusing lens. For our system the associated loss is 7 dB, which is dominated by the mismatch between the 12.7 mm diameter free-space beam and the 7.6 mm clear aperture focusing lens used in front of the pinhole. Future studies that utilize nanophotonic waveguides for DCS may wish to implement tapered regions that more strongly suppress higher-order modes. Longer wavelengths may also be less affected.

After spatial filtering, the DCS spectrum in Figure 2b exhibits remaining structure from the comb spectra as well as a 26 GHz etalon fringe originating from the 2.8 mm long SiN waveguides. This structure can be suppressed by normalizing the measurement through the cell with a background reference spectrum acquired by bypassing the cell via flip-mirrors.

4. Dual-comb spectroscopy of CO and $CO_2$

To characterize the performance of the optimized system, we examine transitions of CO (2 ← 0) and $CO_2$ (20011/20012/20013 ← 00001) in a 0.75 m long single-pass cell filled with 5.6% $CO_2$, 1.7% CO and 92.7% air to 840 mbar. After removing a remaining slowly varying residual spectral variation with a $4^{th}$ order polynomial, we fit the CO and the $CO_2$ bands from the DCS measurement to a model based on Hitran 2016 [32], see Figure 3a. The white noise floor increases towards the edges of the normalized spectrum because of the reduced light intensity. The difference between fitted model and measurement is mostly flat with a standard deviation of 0.01 (as taken over the whole fitted spectral range), showing good agreement between measurement and model. An expanded view in Figures 3b and 3c compares model and measurement for individual CO and $CO_2$ lines.

5. Noise discussion

The peak spectral signal-to-noise ratio (SNR) is 48/√s for the 160 MHz tooth-resolved spectrum. However, for broadband spectral sources like frequency combs, the bandwidth and number of resolved points are equally important metrics. Following ref. [33], we calculate a figure of merit defined as the product of the number of resolved spectral elements and average spectral SNR, normalized by the square root of the acquisition time. The entire DCS spectrum covers ~ 40 THz with 250,000 spectral elements and has an average SNR of 190 for a 60-second-long acquisition, resulting in a figure of merit of $6\times10^6$ /√s, which is comparable to previously reported all-fiber HNLF-broadened DCS in the near-infrared, which has a figure of merit of $7\times10^6$ /√s. [27] This high SNR demonstrates that the supercontinuum generation process in the SiN waveguides does not significantly degrade the overall comb coherence or DCS performance.

Figure 4a compares the relative intensity noise (RIN) of the amplified comb light and the supercontinuum. Even with the significant spectral broadening the supercontinuum light (blue trace) has only slightly higher RIN than the input fiber comb pulses (green trace) and reaches a noise floor of ~ -146 dBc/Hz. In comparison, similar efforts in fiber [34] suffered larger increases in RIN (5-10 dB) despite only broadening to 2.2 µm. Figure 4b shows the SNR of the time-domain interferograms as function of optical power levels on the photodiode. At optical powers above 300 µW the SNR becomes RIN limited for our system.

6. Conclusion

To conclude, we have demonstrated dual-comb spectroscopy using supercontinua generated in SiN-waveguides seeded by compact $Er^+$:fiber combs. After spatial filtering with a pinhole, a smooth spectral shape is obtained. In future systems, improved waveguide designs, in particular different adiabatic coupling tapers for the input and output light, should obviate the need for a pinhole leading to an even more compact, robust system. High spectral fidelity is confirmed by the excellent agreement between measured CO and $CO_2$ line shapes and the HITRAN model. The high SNR of the acquired DCS spectra show that the coherence of the underlying $Er^+$:fiber combs is retained, and the RIN is not significantly increased in the broadening process. The achieved figure of merit of $6\times10^6$ /√s is comparable to all-fiber systems. Here, we targeted the spectral region from 2 µm to 2.5 µm, an important atmospheric window that is not accessible with silica-based fiber systems. Looking forward, the same configuration could be used to access other spectral regions through the near-infrared and potentially into the mid-infrared.

Acknowledgements: We acknowledge funding from the DARPA SCOUT and DODOS programs.  We thank Nima Nader and Henry Timmers for helpful comments on this manuscript, and Su-Peng Yu and Hojoong Jung for advice regarding the SiN chips

# References

1. F. Adler, M.J. Thorpe, K.C. Cossel, and J. Ye, "Cavity-Enhanced Direct Frequency Comb Spectroscopy: Technology and Applications," Ann. Rev. Anal. Chem. **3**, 175 (2010).

2. A. Schliesser, N. Picqué, and T.W. Hänsch, "Mid-infrared frequency combs," Nature Photon. **6**, 440 (2012).

3. I. Coddington, N. Newbury, and W. Swann, "Dual-comb spectroscopy," Optica. **3**, 414 (2016).

4. K.C. Cossel, E.M. Waxman, I.A. Finneran, G.A. Blake, J. Ye, and N.R. Newbury, "Gas-phase broadband spectroscopy using active sources: progress, status, and applications [Invited]," Journal of the Optical Society of America B. **34**, 104 (2017).

5. T. Ideguchi, "Dual-Comb Spectroscopy," Optics & Photonics News. **28**, 32 (2017).

6. A.V. Muraviev, V.O. Smolski, Z.E. Loparo, and K.L. Vodopyanov, "Massively parallel sensing of trace molecules and their isotopologues with broadband subharmonic mid-infrared frequency combs," Nature Photonics. **12**, 209 (2018).

7. S. Okubo, K. Iwakuni, H. Inaba, K. Hosaka, A. Onae, H. Sasada, and F.-L. Hong, "Ultra-broadband dual-comb spectroscopy across 1.0–1.9 µm," Appl. Phys. Express. **8**, 082402 (2015).

8. A.S. Kowligy, A. Lind, D.D. Hickstein, D.R. Carlson, H. Timmers, N. Nader, F.C. Cruz, G. Ycas, S.B. Papp, and S.A. Diddams, "Mid-infrared frequency comb generation via cascaded quadratic nonlinearities in quasi-phase-matched waveguides," ArXiv:1801.07850 [Physics]. (2018).

9. M.A. Foster, A.C. Turner, M. Lipson, and A.L. Gaeta, "Nonlinear optics in photonic nanowires," Optics Express. **16**, 1300 (2008).

10. D.D. Hickstein, H. Jung, D.R. Carlson, A. Lind, I. Coddington, K. Srinivasan, G.G. Ycas, D.C. Cole, A. Kowligy, C. Fredrick, S. Droste, E.S. Lamb, N.R. Newbury, H.X. Tang, S.A. Diddams, and S.B. Papp, "Ultrabroadband Supercontinuum Generation and Frequency-Comb Stabilization Using On-Chip Waveguides with Both Cubic and Quadratic Nonlinearities," Phys. Rev. Applied. **8**, 014025 (2017).

11. D.Y. Oh, K.Y. Yang, C. Fredrick, G. Ycas, S.A. Diddams, and K.J. Vahala, "Coherent ultra-violet to near-infrared generation in silica ridge waveguides," Nature Communications. **8**, 13922 (2017).

12. B. Kuyken, T. Ideguchi, S. Holzner, M. Yan, T.W. Hänsch, J. Van Campenhout, P. Verheyen, S. Coen, F. Leo, R. Baets, G. Roelkens, and N. Picqué, "An octave-spanning mid-infrared frequency comb generated in a silicon nanophotonic wire waveguide," Nat Commun. **6**, 6310 (2015).

13. N. Singh, D.D. Hudson, Y. Yu, C. Grillet, S.D. Jackson, A. Casas-Bedoya, A. Read, P. Atanackovic, S.G. Duval, S. Palomba, B. Luther-Davies, S. Madden, D.J. Moss, and B.J. Eggleton, "Midinfrared supercontinuum generation from 2 to 6 µm in a silicon nanowire," Optica. **2**, 797 (2015).

14. A.R. Johnson, A.S. Mayer, A. Klenner, K. Luke, E.S. Lamb, M.R.E. Lamont, C. Joshi, Y. Okawachi, F.W. Wise, M. Lipson, U. Keller, and A.L. Gaeta, "Octave-spanning coherent supercontinuum generation in a silicon nitride waveguide," Optics Letters. **40**, 5117 (2015).

15. J.W. Choi, G.F.R. Chen, D.K.T. Ng, K.J.A. Ooi, and D.T.H. Tan, "Wideband nonlinear spectral broadening in ultra-short ultra - silicon rich nitride waveguides," Scientific Reports. **6**, 27120 (2016).

16. M.A.G. Porcel, F. Schepers, J.P. Epping, T. Hellwig, M. Hoekman, R.G. Heideman, P.J.M. van der Slot, C.J. Lee, R. Schmidt, R. Bratschitsch, C. Fallnich, and K.-J. Boller, "Two-octave spanning supercontinuum generation in stoichiometric silicon nitride waveguides pumped at telecom wavelengths," Opt. Express, OE. **25**, 1542 (2017).

17. D.R. Carlson, D.D. Hickstein, A. Lind, S. Droste, D. Westly, N. Nader, I. Coddington, N.R. Newbury, K. Srinivasan, S.A. Diddams, and S.B. Papp, "Self-referenced frequency combs using high-efficiency silicon-nitride waveguides," Optics Letters. **42**, 2314 (2017).

18. D. Grassani, E. Tagkoudi, H. Guo, C. Herkommer, T.J. Kippenberg, and C.-S. Brès, "Highly efficient 4 micron light generation through fs-fiber laser driven supercontinuum in Si3N4 waveguides," ArXiv:1806.06633 [Physics]. (2018).

19. H. Guo, C. Herkommer, A. Billat, D. Grassani, C. Zhang, M.H.P. Pfeiffer, W. Weng, C.-S. Brès, and T.J. Kippenberg, "Mid-infrared frequency comb via coherent dispersive wave generation in silicon nitride nanophotonic waveguides," Nature Photonics. 1 (2018).

20. A. Klenner, A.S. Mayer, A.R. Johnson, K. Luke, M.R.E. Lamont, Y. Okawachi, M. Lipson, A.L. Gaeta, and U. Keller, "Gigahertz frequency comb offset stabilization based on supercontinuum generation in silicon nitride waveguides," Opt. Express, OE. **24**, 11043 (2016).

21. D.T.H. Tan, K.J.A. Ooi, and D.K.T. Ng, "Nonlinear optics on silicon-rich nitride—a high nonlinear figure of merit CMOS platform [Invited]," Photonics Research. **6**, B50 (2018).

22. M.H.P. Pfeiffer, A. Kordts, V. Brasch, M. Zervas, M. Geiselmann, J.D. Jost, and T.J. Kippenberg, "Photonic Damascene process for integrated highQ microresonator based nonlinear photonics," Optica, **3**, 20 (2016).

23. N. Nader, D.L. Maser, F.C. Cruz, A. Kowligy, H. Timmers, J. Chiles, C. Fredrick, D.A. Westly, S.W. Nam, R.P. Mirin, J.M. Shainline, and S. Diddams, "Versatile silicon-waveguide supercontinuum for coherent mid-infrared spectroscopy," APL Photonics. **3**, 036102 (2018).

24. H. Bovensmann, J.P. Burrows, M. Buchwitz, J. Frerick, S. Noël, V.V. Rozanov, K.V. Chance, and A.P.H. Goede, "SCIAMACHY: Mission Objectives and Measurement Modes," J. Atmos. Sci. **56**, 127 (1999).

25. G. Ehret, C. Kiemle, M. Wirth, A. Amediek, A. Fix, and S. Houweling, "Space-borne remote sensing of CO2, CH4, and N2O by integrated path differential absorption lidar: a sensitivity analysis," Appl. Phys. B. **90**, 593 (2008).

26. B. Bernhardt, E. Sorokin, P. Jacquet, R. Thon, T. Becker, I.T. Sorokina, N. Picqué, and T.W. Hänsch, "Mid-infrared dual-comb spectroscopy with 2.4 μm Cr2+:ZnSe femtosecond lasers," Appl. Phys. B. **100**, 3 (2010).

27. G.-W. Truong, E.M. Waxman, K.C. Cossel, E. Baumann, A. Klose, F.R. Giorgetta, W.C. Swann, N.R. Newbury, and I. Coddington, "Accurate frequency referencing for fieldable dual-comb spectroscopy," Opt. Express, OE. **24**, 30495 (2016).

28. L.C. Sinclair, J.-D. Deschênes, L. Sonderhouse, W.C. Swann, I.H. Khader, E. Baumann, N.R. Newbury, and I. Coddington, "Invited Article: A compact optically coherent fiber frequency comb," Review of Scientific Instruments. **86**, 081301 (2015).

29. S. Coburn, C.B. Alden, R. Wright, K. Cossel, E. Baumann, G.-W. Truong, F. Giorgetta, C. Sweeney, N.R. Newbury, K. Prasad, I. Coddington, and G.B. Rieker, "Regional trace-gas source attribution using a field-deployed dual frequency comb spectrometer," Optica. **5**, 320 (2018).

30. Disclaimer, "The use of tradenames in this manuscript is necessary to specify experimental results and does not imply endorsement by the National Institute of Standards and Technology," (n.d.).

31. G. Ycas, F.R. Giorgetta, E. Baumann, I. Coddington, D. Herman, S.A. Diddams, and N.R. Newbury, "High-coherence mid-infrared dual-comb spectroscopy spanning 2.6 to 5.2 µm," Nature Photonics. **12**, 202 (2018).

32. I.E. Gordon, L.S. Rothman, C. Hill, R.V. Kochanov, Y. Tan, P.F. Bernath, M. Birk, V. Boudon, A. Campargue, K.V. Chance, B.J. Drouin, J.-M. Flaud, R.R. Gamache, J.T. Hodges, D. Jacquemart, V.I. Perevalov, A. Perrin, K.P. Shine, M.-A.H. Smith, J. Tennyson, G.C. Toon, H. Tran, V.G. Tyuterev, A. Barbe, A.G. Császár, V.M. Devi, T. Furtenbacher, J.J. Harrison, J.-M. Hartmann, A. Jolly, T.J. Johnson, T. Karman, I. Kleiner, A.A. Kyuberis, J. Loos, O.M. Lyulin, S.T. Massie, S.N. Mikhailenko, N. Moazzen-Ahmadi, H.S.P. Müller, O.V. Naumenko, A.V. Nikitin, O.L. Polyansky, M. Rey, M. Rotger, S.W. Sharpe, K. Sung, E. Starikova, S.A. Tashkun, J.V. Auwera, G. Wagner, J. Wilzewski, P. Wcisło, S. Yu, and E.J. Zak, "The HITRAN2016 molecular spectroscopic database," Journal of Quantitative Spectroscopy and Radiative Transfer. **203**, 3 (2017).

33. N.R. Newbury, I. Coddington, and W.C. Swann, "Sensitivity of coherent dual-comb spectroscopy," Optics Express. **18**, 7929 (2010).

34. A. Klose, G. Ycas, D.L. Maser, and S.A. Diddams, "Tunable, stable source of femtosecond pulses near 2 µm via supercontinuum of an Erbium mode-locked laser," Optics Express. **22**, 28400 (2014).

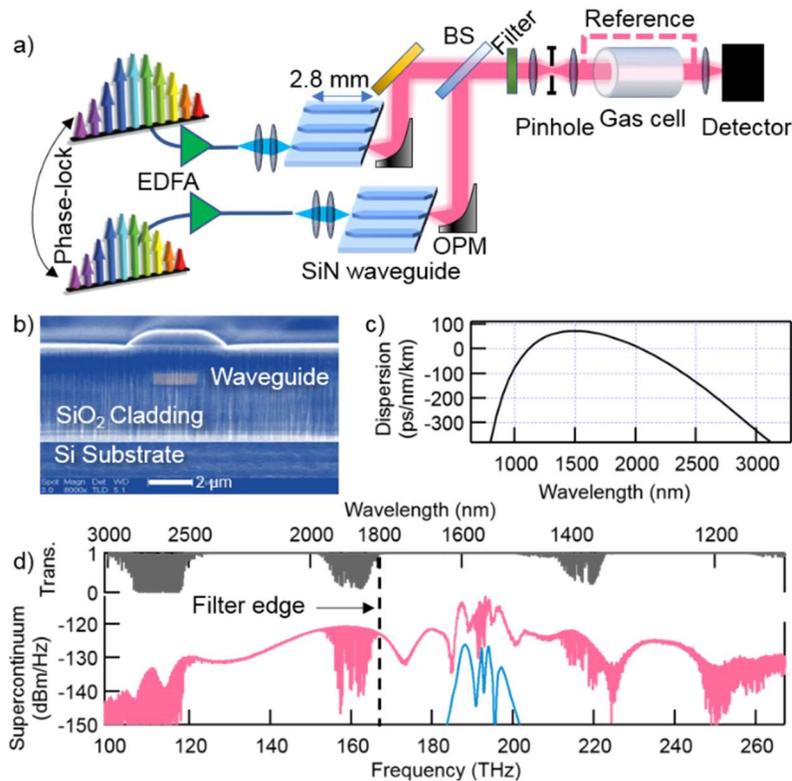

**Figure 1:** a) Experimental setup. Two mutually coherent, self-referenced fiber laser frequency combs are amplified in an Er+ doped fiber amplifier (EDFA) and then focused into the waveguides with aspheric lenses. The supercontinuum output is collimated by an off-axis parabolic mirror (OPM), combined on a $CaF_2$ beam splitter (BS), spatially filtered with a 10 μm diameter pinhole, and finally transmitted through a 0.75 m gas cell filled with $CO_2$ and CO. A reference background spectrum is acquired over the labelled reference path sequentially after bypassing the cell via flip-mirrors. b) Scanning electron microscopy cross section of a non-tapered waveguide facet suspended in $SiO_2$ showing the waveguide dimensions. In the experiment, tapered waveguides are used for increased coupling. c) Calculated waveguide dispersion. d) SiN generated supercontinuum (pink) and the input spectrum (blue), measured with an FTIR, as well as modelled transmission across a 2 m open-air path at 0.5% water (grey).

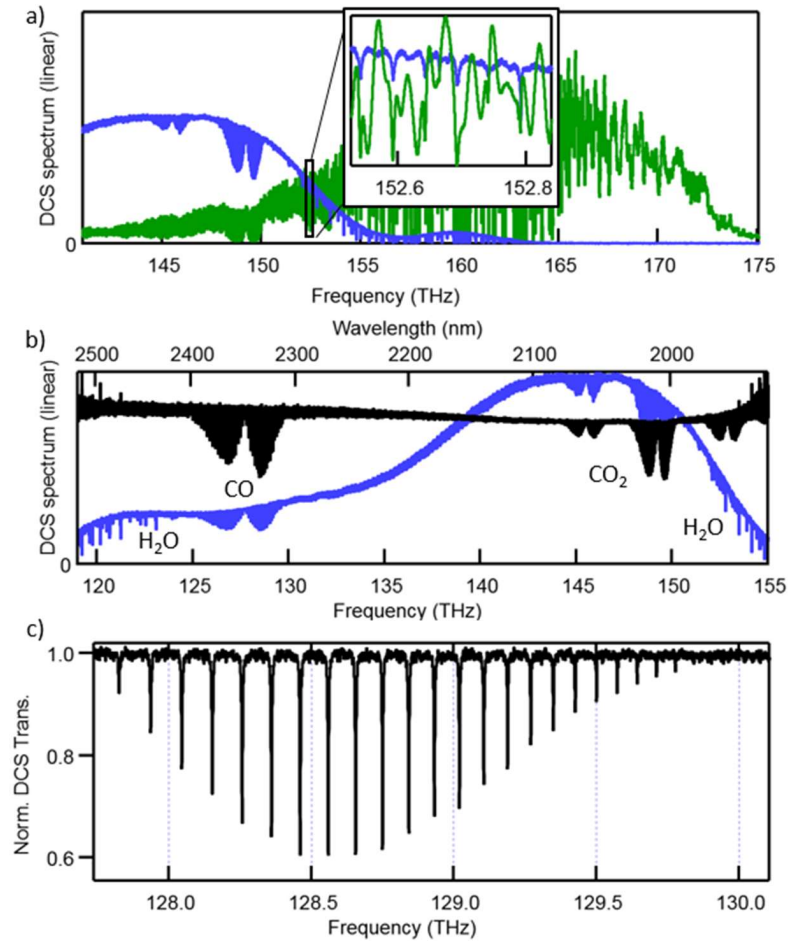

**Figure 2:** DCS spectra (1-minute long measurements, 160 MHz resolution). a) Without pinhole (green) and with pinhole preventing multimode interference (blue), the spectral coverage between the measurements varied due to different input power and input coupling. The narrow absorption features arise from $CO_2$ and water. The inset shows part of the spectrum where the signal-to-noise ratio (SNR) for the two measurements was comparable. b) DCS spectra after spatial filtering with a pinhole (blue trace) and after additional normalization with reference spectrum (black trace). $CO_2$ and CO absorption features from the gas cell are clearly visible. c) Detail of the CO band (smoothed with $10^{th}$ order binomial filter to 1.3 GHz resolution).

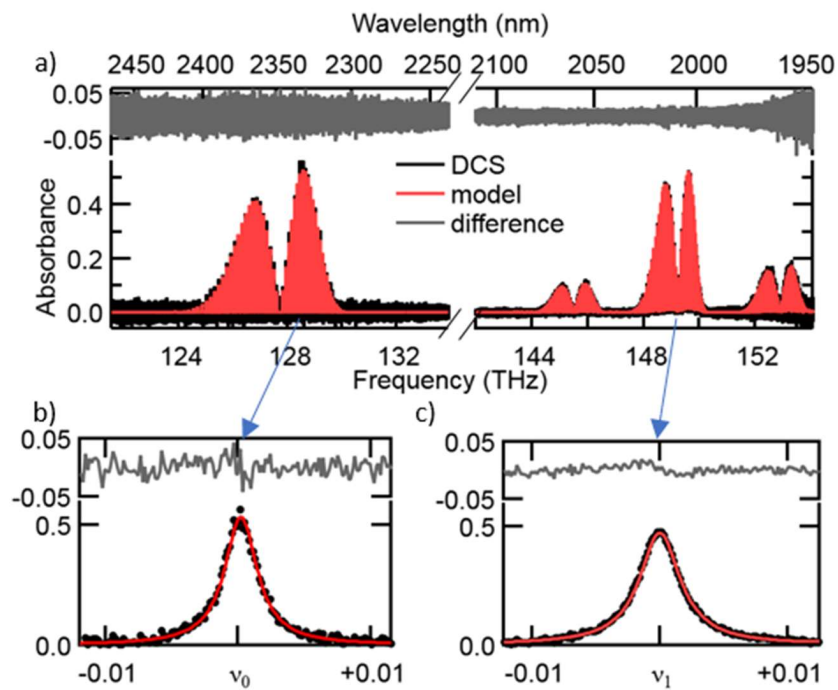

**Figure 3:** Comparison between measurement and model at the combs' native 160 MHz resolution: a) Molecular absorbance measured (black) and from a fitted model based on Hitran 2016 (red), along with their difference (grey). Expanded view of b) a CO absorption line at $\nu_0 = 128.559400$ THz and c) of a $CO_2$ line at $\nu_1 = 148.830794$ THz.

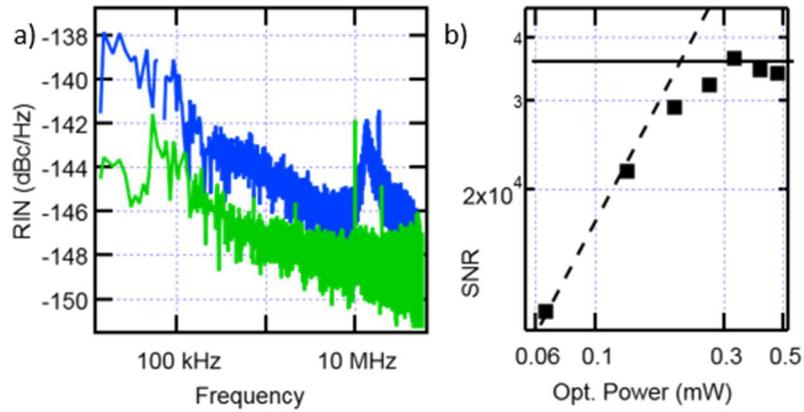

**Figure 4:** a) Relative intensity noise (RIN) measurements of the amplified comb light (green) and the SiN-waveguide broadened supercontinuum after the 1.8 μm filter (blue). (The distinct hump at 13 MHz is typical of this comb design.) b) SNR of the time signal (interferogram) as function of optical power on the detector. Dashed line is the expected detector-noise given a $10\ \text{pW}/\sqrt{\text{Hz}}$ noise-equivalent-power, solid line represents the RIN limit.